# Thermodynamics of Hydrogels for Applications to Atmospheric Water Harvesting, Evaporation, and Desalination

Gang Chen[*]

Department of Mechanical Engineering
Massachusetts Institute of Technology

Most thermodynamic modeling of hydrogels is built on Flory's theories for the entropy of mixing and rubber elasticity, and Donnan's equilibrium conditions if polyelectrolyte polymer and mobile ions are involved. The entropy of mixing depends on number of the solvent and the polymer molecules while the configurational entropy depends on the volume the polymer occupied. Flory's theory treated these two entropic terms in the Gibbs free energy on an equal basis: using the molecular numbers as the variable. I argue that the molecular number and volume are two independent thermodynamic variables, and reformulate Flory's classical hydrogel thermodynamic model by minimizing the Helmholtz free energy of a combined system consisting of the hydrogel and its ambient. This treatment enables us to unequivocally state that the osmotic pressure is the thermodynamic pressure of the solvent inside the hydrogel and to unambiguously write down the chemical potential of each species. The balance of the chemical potentials of the mobile species, including both the solvent and the mobile ions gives a set of equations that can be simultaneously used to solve for the equilibrium volume of the hydrogel, the osmotic pressure, and the Donnan potential, including their coupling. The model is used to study thermodynamic properties of both pure and salty water in non-electrolyte and electrolyte hydrogels such as (1) the latent heat of evaporation, (2) the ability of hydrogels to retain water and to absorb water from the atmosphere, (3) the use of hydrogels for desalination via solar or forward osmosis, (4) the antifouling characteristics of hydrogels, and (5) melting point suppression and boiling point elevation, and solubility of salts in hydrogels. These properties are of interests in solar-driven interfacial water evaporation for desalination and wastewater treatment, atmospheric water harvesting, and forward osmosis. The reformulated thermodynamic framework will also be useful for understanding polymer electrolytes and ion transport in electrochemical and biological systems.

---

[*] Email: gchen2@mit.edu

## 1. Introduction

Hydrogels responsive to different stimuli such as temperature, pH, salt concentration, electrical field, and light have found a wide range of applications in biology, biomedical technology, and environmental technology, and daily life (food, diapers, etc.).[1,2] More recently, potential applications of hydrogel in desalination and atmospheric water harvesting have also drawn increasing attention with the demonstration of apparent reduced latent heat[3,4] during solar-driven interfacial water evaporation,[5,6] hydrogel as a draw solution for forward osmosis,[7,8] and adsorption of water from atmosphere and regeneration of adsorbed water via phase separation with minimal energy input.[9] Current studies are mostly experimental, understanding the thermodynamics underpinning these processes is of great interests to these applications and future advancements of these applications.

Flory and Rehner developed basic thermodynamic framework for analyzing hydrogel swelling,[10,11] which served as the starting point for most subsequent developments. Their theory considered four additional contributions in the Gibbs free energy: the (1) entropy of mixing between the hydrogel polymers and the solvent molecules,[12,13] (2) an enthalpic term due to binding of solvent molecules with the polymers, (3) an elastic energy term due to the configurational entropy change of the crosslinked polymer chains,[14] and (4) for polyelectrolyte hydrogels, an ionic osmotic pressure term. Katchalsky and Michaeli[15] further included the electrostatic interactions between charged polymers of polyelectrolyte and surrounding ions in the Gibbs free energy based on the Debye-Huckle theory and their theory was later modified considering, for example, finite chain extensibility.[16] English et al.[17] developed a quasi-lattice model to account for charge on the polyelectrolyte and modified the Flory-Rehner theory correspondingly to model polyelectrolyte hydrogels. Brannon-Peppas and Peppas[18] extended the Flory-Rehner theory to include the wet initial state of the hydrogel and the dissociation of ions from weak polyelectrolyte polymers. In analyzing hydrogel's use in a power generation cycle based on the salinity gradient, Zhang et al.[19] extended the previous thermodynamic models to include the effects of salts and fixed charge on polymer. In general, ionization of weak polyelectrolyte is very complicated, as many studies based on modeling and simulations have shown.[16, 20,21,22,23] Hong et al.[24] took a different approach, developing a field theory for hydrogel through analyzing the Helmholtz free energy under constraints at boundaries. In their derivation, the osmotic pressure emerges from a Lagrangian multiplier to the Helmholtz free energy due to the volume constraint, but its exact value changes with different choices of variables in the Helmholtz free energy.

Most applications of these thermodynamic models focused on the gel swelling ratio. For applications such as desalination,[7,8] atmospheric water harvesting,[9, 25] solar-driven evaporation,[3,4,5,6] the thermodynamics of water in hydrogel is of greater interests. Some experimental observations await explanations and call for deeper understanding of water state inside hydrogel from thermodynamics perspective. For example, can water have reduced latent heat in hydrogel? And if yes, by how much?[3] What kind of hydrogels can serve as draw solutions for forward osmosis?[7] Why do not hydrogels get fouled under solar-driven interfacial evaporation of saline water?[26] Why can adding salt prevent hydrogels from drying up?[27] Is there a thermodynamic basis for the classification of water in hydrogel into: bound, intermediate, and bulk water?[28,29] Although thermodynamic analysis of water may not answer all these questions, especially considering how complex water is even in its pure bulk form,[30,31] it will be interesting to examine what can be learnt from studying thermodynamics of water in hydrogel.

In my attempt applying current models to explore answers to the above equations, I found that there are some flaws and ambiguities in the established classical approaches. For example, Donnan equilibrium[32] is used to find out the concentrations of different ions inside and outside the gel, which are then used to compute part of the osmotic pressure. However, the classical Donnan equilibrium conditions neglected the pressure effect on ion concentrations,[37] whose inclusion effectively couples the electrostatic potential and the osmotic pressure. It also seems to me that Flory's treatment of osmotic pressure term is a little bit ad-hoc, although his results are self-consistent. As I will discuss in detail, Flory did not clearly distinguish in his theory what are proper thermodynamic variables. Such ambiguities create confusion in literature. For example, Lai et al.[33] wrote "In the literature, however, it is less than clear, how swelling pressure, Donnan osmotic pressure or solid matrix stress are related to one another. For example, is the swelling pressure the same as the Donnan osmotic pressure for cartilage?".

This paper has two objectives. The main goal is to reformulate Flory's thermodynamic model. The reformulation is based on minimizing the Helmholtz free energy of the combined system consisting of the hydrogel and its ambient. This reformulation clearly shows that the thermodynamic pressure of the solvent inside the pressure is the osmotic pressure, which is at equilibrium with the tension from the polymer. The formulation allows unambiguous calculation of the chemical potentials of each species. At equilibrium, the chemical potential balance of mobile species inside and outside hydrogels determines the equilibrium volume, the osmotic pressure, and the Donnan potential. My other goal is to apply the reformulated thermodynamic analysis to water in hydrogels to gain better insights related to hydrogel applications in solar-driven interfacial evaporation, atomospheric water harvesting, and forward osmosis desalination, and we will use the reformulated thermodynamic framework to exam these problems. We will start with a brief review (Sec.2) of Flory's classical model and some subsequent modifications of the model, and point out existing confusions and deficiencies. Modifications to the models, for both non-electrolyte and polyelectrolyte hydrogels will be presented (Sec.3) and applied to analyze water thermodynamic properties in hydrogels, including the latent heat of evaporation, atmosphere water absorption and water retention ability, antifouling characteristics, melting point suppression and boiling point elevation (Secs.4-6). Although the models on free energy and entropy terms I used are at the basic level Flory used, following which many other studies had been done, I believe that the reformulated thermodynamic approach is general and can be expanded readily to include more accurate descriptions of the polymers, the ions, and the solvent. The thermodynamic framework should be useful for related work in biological and biomedical applications for which osmosis and Donnan membrane equilibrium are widely used concepts.

## 2. Review of Flory's Theories for Hydrogel Swelling

Flory's thermodynamic model is based on his derivations of the entropy between a solvent and a polymer and the configurational entropy for the polymer elasticity (see Supplementary Information (SI) ).[10,11,12] Using these expression, the change of the Gibbs free energy (the final state minuses the initial state), $\Delta G$, of the polymer-solvent system can be expressed as

$$\Delta G = k_B T(n_1 ln f_1 + n_2 ln f_2) + k_B T \chi n_1 f_2 + \frac{k_B v_e T}{2}(3a_s^2 - 3 - ln a_s^3) \tag{1}$$

with

$$f_1 = \frac{n_1}{n_1 + Zn_2}, \quad f_2 = \frac{Zn_2}{n_1 + Zn_2} \quad a_s^3 = \frac{V}{V_o} = \frac{V_o + n_1 v_1}{V_o} = \frac{1}{f_2}$$

$$v_e = v - 2n_2 = v(1 - 2n_2/v) = v(1 - 2M_c/M) \tag{2}$$

where $n_1$ is number of molecules of the solvent and $n_2$ the number of the polymer molecules, $f_1$ and $f_2$ are the volume fraction of the solvent and the polymer, respectively, $M_c$ and $M$ are the molecular weight of a crosslinked segment and the polymer molecular before crosslinking, respectively, $v$ is the number of the actual cross-linking units and $v_e$ excludes the two free ends of the original polymer molecules, $a_s$ is the ratio of the chain extension in one-direction for isotropic swelling which can be related to the volume after (V) and before ($V_o$) swelling as given, $v_1$ is the volume of a solvent molecule, and Z is the volume ratio of the polymer molecule to the solvent molecule. The first term in Eq. (1) is due to entropy of mixing between the solvent and the polymer, and the last term due to the configuration entropy change of the polymer, and middle term due to enthalpy change, with $\chi$ represents the polymer-solvent interaction also called Flory-Huggins parameter, which depends in general on both temperature and polymer volume fraction. $\chi>0$ means the mixing is endothermic and $\chi<0$ exothermic. Most polymer-solvent mixing has $\chi>0$. We note here that although the middle term is called the enthalpy change, no pressure effect was considered in the derivation of the above expression.[12,34] Inside hydrogels, the pressure is usually much higher than the ambient pressure. This pressure effect is not included in the expression. Also note that this treatment does not distinguish three different types of waters.[35,36]

The change in the chemical potential for the solvent can be obtained from the change in the Gibbs free energy by $N_A \left(\frac{\partial \Delta G}{\partial n_1}\right)_{T,p,n_2}$ on a molar basis, where $N_A$ is the Avogadro number,

$$\mu_1 - \mu_1^o = (\Delta \mu_1)_{pl} + (\Delta \mu_1)_{el} = RT\left[ln(1-f_2) + \left(1-\frac{1}{Z}\right)f_2 + \chi f_2^2\right] + \frac{RT v_w n_e}{V_o}\left(f_2^{1/3} - f_2/2\right) \tag{3}$$

where $n_e = v_e/N_A$ is the molar number of the effective crosslinking unit, R(=$k_B N_A$) the universal gas constant, and $v_w$ the molar volume. Note that in deriving the above expression from Eq. (1), the volume variable V is substituted by its relationship with the molecular number as given in Eq. (2), and also note that Eq. (1) does not have pressure as an independent variable, although such a variable is expected for the Gibbs free energy. The first term in the above expression is due to the solvent-polymer interaction and $-(\Delta \mu_1)_{pl}/v_w$ was identified by Flory as the osmotic pressure. The second term is due to the polymer conformation change and $(\Delta \mu_1)_{el}/v_w$ was identified as the stress on the polymer.

For polyelectrolyte hydrogels, Flory considered a strong polyelectrolyte that dissociates completely, yielding mobile anions $A^{z-}$, where the superscript $z_-$ denotes the valence of the ion (and hence its charge is $ez_-$), and fixed positive counterions $ie$ per structural unit of the polymer ("i" is the valence of a monomer). The gel is in equilibrium with the surrounding solvent containing a strong electrolyte $M_{\nu_+}A_{\nu_-}$ ($z_+\nu_+ = z_-\nu_-$) at concentration $c_s^*$. Since the gel initially contains no $M^{z+}$ cations, some of them will diffuse into gel and some $A^{z-}$ will diffuse out. Because charge neutrality conditions should satisfy both inside and outside the gel, if $c_s$ [mole/m³] is the concentration of the $M_{\nu_+}A_{\nu_-}$ salt inside the gel, the concentrations of mobile ions in the gel are $c_+ = \nu_+ c_s$ and $c_- = \nu_- c_s + ic_2/z_-$, where $c_2$ is the concentration of the fixed charge per unit volume, $c_2 = f_2/v_u$, and $v_u$ is the molar volume of a structural unit. The total mobile ion concentration ($c_+$+$c_-$) inside the gel at equilibrium will inevitably exceed that in the external solution $c_+^*+c_-^*=\nu c_s^*$, where $\nu=\nu_++\nu_-$. The osmotic pressure arising from the difference of the mobile ion concentration is

$$\pi_i = RT(c_+ + c_- - c_+^* - c_-^*) = RT\left[\frac{ic_2}{z_-} - \nu(c_s^* - c_s)\right] \tag{4}$$

To determine $c_s^*-c_s$, Flory used typical the Donnan equilibrium condition,[32]

$$a_+^{\nu+}a_1^{\nu-} = a_+^{*\nu+}a_1^{*\nu-} \tag{5}$$

where a is the activity of the perspective ions. Flory further assumed that the activity coefficient equaling one. Brannon-Peppas and Peppas[18] extended the above treatment to weak polyelectrolytes.

### 3. Reformulation of Flory's Hydrogel Thermodynamic Model

There are some confusions in the Flory hydrogel model which persisted throughout subsequent developments. I have mentioned some: (1) the Gibbs free energy expression in Eq. (1) does not include pressure as its natural thermodynamic variable, (2) the volume variable in the configurational entropy is replaced by the molecule number variable, and (3) the enthalpy of mixing term does not explicitly include the pressure effect. Overall, the effect of pressure was included in ad-hoc ways, leading to confusions as reflected in the question asked by Lai et al.[33] quoted before. We will show below that the first and the last term in Eq. (1) should not be treated on an equal basis because they are governed by different natural thermodynamic variables: the molecular number for the entropy of mixing and the volume for the configurational entropy. Flory's theory neglected the electrostatic potential of the ions on the polyelectrolyte and interactions between the mobile ions and the polyelectrolyte, which subsequent work tried to include.[17,19] These treatments, however, neglected the coupling between the Donnan potential and the osmotic pressure, a situation persisted from Donnan's original treatment on the subject.[32,37] In the following, I will present a reformulation of Flory's theory addressing the concerns raised here.

A thermodynamic analysis typically starts with defining a system in contact with its environment at specified conditions. When the system is at equilibrium with the environment, its thermodynamic potential either maximizes or minimizes. For a system at constant temperature and pressure in equilibrium with its environment that it can exchange heat, force, and particle, its thermodynamic potential, the Gibbs free energy, is minimized and the natural thermodynamic variables are temperature and pressure. A system maintained at constant temperature and volume minimizes the Helmholtz free energy. For a hydrogel in equilibrium with its solvent, the problem is that its volume is not constrained, nor it is at the same pressure with the ambient (due to osmotic pressure). We thus consider a combined system made of the gel and the ambient solvent as a closed system at constant temperature and volume. At equilibrium, the total Helmholtz free energy of this combined system is at its minimum.

The Helmholtz free energy, U-TS, of the hydrogel sub-system can be written as

$$F = U_{sl} + U_{pl} + U_{ion} + U_{mix} + U_{Coul} - TS_{ion} - TS_{sl} - T(S_{pl} + \Delta S_{pl}) - TS_{mix}$$

$$= (U_c - TS_c) + U_{mix} - TS_{mix} - T\Delta S_{pl} = F_c + \Delta F_{mix} + \Delta F_{pl} \quad (6)$$

where $U_{sl}$ and $U_{pl}$ are the internal energy of the solvent inside the gel and the polymer, respectively, $U_{Coul}$ is the electrostatic energy while $U_{ion}$ is the internal energy of the ions, $U_{mix}$ is the change of internal energy due to mixing, which is essentially second term in Eq. (1), despite that it is called enthalpy in the past literatures. Note that we split $U_{Coul}$, the Coulomb energy of ions and polymers, from the rest of the internal energy of the ions $U_{ion}$ and the polymer $U_{pl}$. $S_{sl}$ is the entropy of the solvent, $S_{pl}$ the entropy of the polymer in its initial state and $\Delta S_{pl}$ and entropy change of the polymer network due to its configurational change, and $S_{mix}$ the entropy of mixing. $U_c$, $S_c$, and $F_c$ denote the internal energy, the entropy, and Helmholtz free energy of the solution (including ions) imbibed into the polymer plus that of the original polymer. The last two terms in the third equality of Eq. (6) are extra: $\Delta F_{mix} = U_{mix} - TS_{mix}$ is the increase in the Helmholtz free energy caused by mixing, and $\Delta F_{pl} = -T\Delta S_{pl}$ is caused by the change of the configurational entropy of the polymer due to deformation, assuming that polymers' own internal energy remain unchanged. Later work had included finite extensibility of the polymers, and the corresponding contribution can be included into $\Delta F_{pl}$.[16] In Flory's theory, the first two terms in Eq. (1) made up $\Delta F_{mix}$ and the last term is $\Delta F_{pl}$, which we will use for simplicity.

The natural variables for the Helmholtz free energy are T, $V_t$, φ, $n_i$, where φ is the electrostatic potential, $n_i$ the number of species i, and $V_t$ total volume that is the sum of hydrogel volume V and the volume of the ambient solvent $V_e$. The change in the total Helmholtz free energy of the combined system can be written as (SI)

$$dF_t = \left[ p_e - p + \left(\frac{\partial(\Delta F_{mix})}{\partial V}\right)_{T,\varphi,n_i} - \left[\left(\frac{\partial(\Delta F_{mix})}{\partial V}\right)_e\right]_{T,\varphi,n_i} + \left(\frac{\partial(\Delta F_{pl})}{\partial V}\right)_{T,\varphi,n_i} \right] dV$$

$$+ [\mu_{sl} - \mu_{sl,e}] dn_{sl} + \sum_i [\mu_i - \mu_{i,e}] d- + \sum_i ez_i n_i d\varphi + ez_{pl} n_{pl} d\varphi + \left(\sum_i ez_i n_i d\varphi\right)_e$$

(7)

where $p = -\left(\frac{\partial F_c}{\partial V}\right)_{T,\varphi,n_i}$ is the thermodynamic pressure acting on solvent, ions, and the polymer, $p_e$ is the pressure of the ambient solvent (which could also be extended to include externally applied pressure), $\mu_i = \left(\frac{\partial F}{\partial n_i}\right)_{V,T,\varphi,n_{j \neq i}}$ the chemical potential of species i, and $z_i$ the valence of the ion denoted by the subscript.

At equilibrium, $dF_t = 0$, which leads to

$$p - p_e = \left(\frac{\partial(\Delta F_{pl})}{\partial V}\right)_{T,\varphi,n_i} = -\left(\frac{\partial(T\Delta S_{pl})}{\partial V}\right)_{T,\varphi,n_i} \tag{8}$$

$$\mu_{sl} = \mu_{sl,e}, \quad \mu_i = \mu_{i,e} \tag{9}$$

$$\sum_i z_i n_i + z_{pl} n_{pl} = \left(\sum_i z_i n_i\right)_e = 0 \tag{10}$$

where we have neglected possible volume dependence of $\Delta F_{mix}$. Equation (8) is the condition of mechanical equilibrium. The pressure difference between inside the gel and outside, which is the osmotic pressure, balances the tension created by the polymer configuration change. Equations (9) are conditions for chemical equilibrium, or, mass transfer equilibrium, for the solvent and the mobile ions. And Eqs. (10) are the charge neutrality requirement for both inside and outside.

Flory's expression for the polymer Helmholtz free energy changes due to configurational entropy, i.e., the last term in Eq. (1), can be expressed in terms of volume for isotropic swelling so tthat he the Helmholtz free energy can be written as

$$\Delta F_{pl} = \frac{RT n_e}{2} \left[ 3\left(\frac{V}{V_o}\right)^{2/3} - 3 - \ln\left(\frac{V}{V_o}\right) \right] \tag{11}$$

In hydrogel, solvent fill the empty space between polymer molecules such that $V = V_o + n_1 v_1$. Flory used this relation to further write $\Delta F_{pl}$ in terms of the solvent number $n_1$. By doing so, the polymer tension due to configurational entropy change i.e., the $\frac{\partial \Delta F_{pl}}{\partial V}$ term in Eq.(8), is grouped into the chemical potential term $\frac{\partial \Delta F_{pl}}{\partial n}$, as in Eq.(3). Although Flory correctly pointed out that the mixing term, i.e., the first term in Eq.(3), creates osmotic pressure, and the second term represents the polymer tension, the change of natural variable from V to $n_1$ causes confusion since the polymer tension term becomes part of chemical potential. By using the volume as an independent variable (as in the case of rubber elasticity), we see clearly from Eq. (7) that pressure p is the thermodynamic pressure acting on all components inside the hydrogel. From Eq. (8), we see that the difference of p and $p_e$, i.e., the osmotic pressure, balances the polymer tension created by the polymer configuration change. We will see later that p can be determined from the balance of chemical potentials as represented by Eq. (9), rather than simply the first term in Eq. (3). So, in answering Lai et al.'s question,[33] the swelling pressure and the Donnan osmotic pressure are identical, while the solid matrix stress arises from different source (the polymer configurational entropy change in Flory's model). They balance each other in mechanical equilibrium but can create motion in nonequilibrium situations.

For a non-electrolyte hydrogel with pure solvent, we can write the extra molar-based Helmholtz free energy in Eq. (6) as

$$\Delta F = RT[(n_1 \ln f_1 + n_2 \ln f_2) + \chi n_1 f_2] + \frac{RT n_e}{2}\left[3\left(\frac{V}{V_o}\right)^{2/3} - 3 - \ln\left(\frac{V}{V_o}\right)\right] \qquad (12)$$

which is identical to Eq. (1). What is important, however, is that we recognize now that the first term, arising from the mixing entropy and internal energy, depends on natural thermodynamic variable $n_1$ and $n_2$ (both $f_1$ and $f_2$ can be written in terms of these variables as in Eq.(2)). The second term depends on volume. Using Eq. (8), we get

$$p - p_e = \left(\frac{\partial \Delta F_{pl}}{\partial V}\right)_{T,n_2} = \frac{RT n_e}{2V_o}\left[2\left(\frac{V}{V_o}\right)^{-1/3} - \frac{1}{\left(\frac{V}{V_o}\right)}\right] = \frac{RT n_e}{V_o}\left[f_2^{1/3} - \frac{f_2}{2}\right] \qquad (13)$$

The chemical potential of the solvent is

$$\mu_1 = \left(\frac{\partial F_c}{\partial n_1}\right)_{V,T,n_2} + \left(\frac{\partial \Delta F_{mix}}{\partial n_1}\right)_{V,T,n_2} = \left(\frac{\partial F_{sl}}{\partial n_1}\right)_{V,T,n_2} + \left(\frac{\partial \Delta F_{mix}}{\partial n_1}\right)_{V,T,n_2}$$

$$= \mu_1^*(T,p) + RT\left[\ln(1-f_2) + \left(1-\frac{1}{Z}\right)f_2 + \chi f_2^2\right] \qquad (14)$$

where $\mu_1^*(T,p)$ is the chemical potential of the pure solvent inside the gel at pressure p and temperature T. The 1/Z term in the above equation is often neglected as it is usually much smaller than 1. We will take this approximation in the rest of this paper.

Since the chemical potentials are important for our analysis, we will elaborate a little more. We consider a solution consists of $x_1, .. x_i, .. x_n$ mole fractions of components *1,..., i,..., n*. We start from the definition of the chemical potential of component i on a molar basis[38,39,40]

$$\mu_i(T,p,\varphi,x_1,..x_i,..x_n) = N_A\left(\frac{\partial G}{\partial n_i}\right)_{T,p,\varphi,n_{j(\neq i)}} \qquad (15)$$

The derivative of the chemical potential can be written as

$$d\mu_i(T, p, \varphi, x_1, ..x_i, ..x_n) = v_i dp + z_i F_D d\varphi + RT d\ln(a_i) \qquad (16)$$

where $a_i$ and $v_i$ are the activity and the molar volume of species i, respectively, and $F_D$ the Faraday constant. The activity can be written in terms of the mole fraction using the activity coefficient $\gamma_i$ as $a_i = \gamma_i x_i$, Integrating Eq.(16) and assuming that the molar volume $v_i$ is a constant, we have

$$\mu_i(T, p, \varphi, x_1, ..x_i, ..x_n, x_s) = \mu_i^o(T, p_o) + v_i(p - p_o) + z_i F\varphi + RT\ln(\gamma_i x_i) \qquad (17)$$

where $\mu_i^o(T, p_o)$ is the chemical potential of species i at the reference state (T,$p_o$).

One question is how to include the activity coefficient into the entropy of mixing expression of Flory in Eq. (1) for ionic species. This is necessary if we want to study the impact of different ions, for example, in the hydrogel water retention experiments.[27] For a pure water-hydrogel system, we can get the water activity coefficient from Eqs.(14) and (17)

$$RT(\ln(1 - f_2) + f_2 + \chi f_2^2) = RT\ln(\gamma_w x_w) \qquad (18)$$

where the subscript "w" represents water, which we will assume to be the solvent since we will focus on hydrogel. For regular water-ion solutions, extensive studies exist on the activity coefficients of different ions as a function of the ion concentration and the water activity coefficients.[41, 42] In fact, the activity coefficients of a solution are related via the Gibbs-Duhem relation. Knowing the activity coefficients of ions as a function of concentration, the activity coefficients of water can be calculated.[38]

Ions in hydrogel interact with both water and the polyelectrolyte polymer. Surrounding the charged polymer, ions can be immobilized, forming condensates as pictured in Manning's theory,[43] in addition to the diffuse double layer as in the Debye-Huckle theory[44]. The net effect of the electrostatic interactions of the charges along the chain among themselves and with the mobile ions be expressed in terms of free energy, for which different models have been developed.[15,16,17] Given that the detailed mechanisms of ions in polymeric solutions remain an activity field of study due to their importance in biology and electrochemical technologies, I decide to include only the mixing term as Flory originally assumed, not because this is a better treatment, but is a reflection of my view of the large uncertainties in the current models. Under this approximation, the additional Helmholtz free energy change due to the mixing of ions with polymers can be expressed as

$$\Delta F_{mix} = k_B T \left[ n_w \ln f_w + \sum_i n_i \ln f_i + n_2 \ln f_2 \right] + k_B T \chi \left( n_1 + \sum_i n_i \right) f_2 \qquad (19)$$

where we have treated water and ions as having same volume and same mixing enthalpy parameter $\chi$ with polymer. This mixing term together with the polymer free energy change due to the polymer configurational entropy, Eq. (11), is used to form the total free energy change of polyelectrolyte hydrogel similar to Eq. (12). These assumptions could be eliminated by assuming different χ values and different sizes of ions compared with water, but will add mathematical complexity unnecessarily considering the approximations we already made. From Eq. (19), we can get the excess chemical potential for water and ions

$$\mu_w - \mu_w^*(T,p) = RT[\ln f_w + f_2 + \chi f_2^2] \approx RT[\ln(1-f_2) + f_2 + \chi f_2^2 + \ln(\gamma_w x_w)] \quad (20)$$

$$\mu_i - \mu_i^*(T,p) = RT[\ln f_i + f_2 + \chi f_2^2] \approx RT[\ln(1-f_2) + f_2 + \chi f_2^2 + \ln(\gamma_i x_i)] \quad (21)$$

where $f_w$ and $f_i$ are the volume fraction (based on total volume including that of polymers) of water and the ith ion inside the hydrogel, respective, and $x_w$ and $x_i$ are the mole fraction of water and ions in the water-ion solution without considering polymer's existence, respectively, which means $x_w + \sum_i x_i = 1$. In writing the second step in Eq. (20), we have used the relation

$$f_w = \frac{n_w}{n_w + Zn_2 + \sum_i n_i} = \left(\frac{n_w}{n_w + \sum_i n_i}\right)\left(\frac{n_w + \sum_i n_i}{n_w + Zn_2 + \sum_i n_i}\right) = x_w(1-f_2) \quad (22)$$

and similarly for $f_i$. Also, we added activity coefficients in front of $x_w$ and $x_i$ to account for nonidealities. We will use Eqs. (20) and (21) for chemical potentials of water and ions in the rest of the manuscript, keeping in mind that $\mu_w^*$ and $\mu_i^*$ depend on pressure and electrostatic potential as shown in Eq. (17). Although our numerical calculations will take activity coefficients as one most of the time, we will include examples of different salts by using their actual activity coefficients to illustrate the impacts of nonunitary activity coefficients.

With the above approach, we have reformulated Flory's thermodynamic model under the Helmholtz free energy picture. The reformulation enables us to clearly define the thermodynamic pressure, and relative its difference with the environmental pressure, i.e., the osmotic pressure, to that of the tension in the polymer. With the new formulation, we can express the chemical potentials of each species inside hydrogel as exemplified in Eqs. (20) and (21). These chemical potentials depend on temperature, pressure, electrostatic potential, and composition. We will show next that the chemical potential balance between mobile species inside and outside determines the pressure, volume, composition, and electrostatic potential. We are interested in using the reformulated thermodynamic model to understand some recent experiments on using hydrogel for solar-driven interfacial evaporation, atmospheric water harvesting, and forward osmosis desalination. It is difficult to do quantitative comparison with experimental data because parameters needed for the model are usually unknown. Our focus is thus on the trend rather than trying to be quantitative. We will give below a summary of experimental facts that serve as inspiration for our subsequent discussion.

Solar-driven interfacial evaporation uses sunlight to heat up a floating porous solar absorber to create a hot region on the water surface, which enhances the evaporation of water while the bulk water remains cold.[5,6] This approach had drawn lots of interests for their potential applications in desalination and waste water treatment. The theoretical thermal evaporation limit for standard 1 sun condition is 1.45 kg/m²-hr, assuming the 1000 W/m² solar energy in 1 sun is used to heat up and evaporate water from an ambient temperature of 25 °C to 40 °C. Yu's group[9] used porous polyvinyl alcohol (PVA) hydrogels embedded with black absorbers and reported evaporate rate as high as 4-5 kg/m$^2$-hr, followed by other similar reports using different hydrogels and even none hydrogels.  The explanation for such a phenomenon championed by the Yu group was the existence of three kinds of waters in hydrogel, the bound water, the intermediate water, and the free water.[9,28,29]  Some key evidences supporting this explanations are (1) latent heat reduction measured by differential scanning calorimetry experiments, (2) suppressed melting point of intermediate water up to -20 °C, also observed using DSC measurement.[9,45]  Another key experimental evidence is the Raman spectroscopy signatures of different waters.  In the DSC latent heat measurement, we can also see that the end of the evaporation excursion curve exceeds the boiling point of pure water, which can be understood as the elevation of the boiling point.  However, the maximal measured latent heat reduction with DSC is up to 30%, while experimentally, evaporation rate over 5 kg/m$^2$-hr had been reported, which is about three times of the thermal evaporation limit.  Another interesting observation is that in solar-driven desalination experiment, salt does not seem to accumulate inside hydrogel.[46]  We will show that the higher pressure of water inside hydrogel can lead to reduced latent heat up to the order observed in the DSC experiment, but not to the same order observed in the solar-driven experiment.  The thermodynamic model can also provide qualitative explanations for the experimental observation of freezing point depression and antifouling characteristics of hydrogel.  The results show that the DSC-based latent heat reduction and freezing point suppression do not provide sufficient proof that the experimental observations of higher evaporation rate are due to the existence of the three types of water inside hydrogel.

When a hydrogel is placed in open air, it usually dries up over time.  How to retain water is a challenge.  Bai et al.[27] tested the water retention capability of polyacrylamide (PAM) hydrogel impregnated with different salts such as sodium chloride (NaCl), lithium chloride (LiCl), magnesium chloride (MgCl$_2$) and potassium acetate (KAc).  They found for example, NaCl precipitates out, but LiCl can retain some water for a long time.  While these experiments aim to retain water, adsorbing water from atmosphere is of interests in addressing the water shortage challenge in many regions.  Adsorbents made of PAM hydrogel composite with carbon nanotubes and Ca$_2$Cl[47], and composite hydrogel made of the poly N-isopropylacrylamide (PNIPAM) and chloride-doped polypyrole[9] have shown promise.  These examples also show the importance of ions for water adsorption. Hydrogel's water retention and adsorption capabilities depend on the relative humidity of the ambient air.  We will examine the problem of water retention and atmospheric water extraction using the thermodynamic model, considering the effect of salts in both nonpolyelectrolytic and polyelectrolytic hydrogels.  As we already mentioned before, ions in polyelectrolyte hydrogels are very complex.  In fact, even without hydrogels, ion interaction with water is already complex.  Our modeling did not include such complex interactions, as we aim at seeing the trend rather than being quantitative.

The ion concentrations in polyelectrolyte hydrogel in an electrolytic solution are usually modeled using the method originally developed by Donnan,[32] as represented by Eqs. (4) and (5).  Accompanying the Doannan equilibrium is the Donnan potential, a voltage difference between the hydrogel and the external electrolytic solution.  The Donnan potential distribution across a hydrogel-solvent interface had been measured by Gong's group with a delicate electrolyte technique.[48] I had shown in a previous paper[37] that the traditional Donnan equilibrium neglected the coupling between the osmotic pressure and the Donnan potential. I will show using the reformulated hydrogel thermodynamic framework that the inclusion of this coupling leads an additional potential which would not exist for a nonelectrolyte hydrogel immersed in a salt solution. There are no published results confirming this prediction. I learnt from private communication with members from Profesor Gong's group that they had observed potentials drop in PVA hydrogels immersed in salt solutions.[49]

The reformulated thermodynamic model also makes it easy to calculate osmotic pressure which is relevant to the work in using hydrogels for desalination based on forward osmosis. In this technology, hydrogels with no or little water at the beginning but with a high osmotic pressure when at equilibrium is used to extract clean water from saline water separated with a semipermeable memberane.[7,8] The osmotic pressure difference between the saline water and water in hydrogel drew clean water across the member to expand hydrogel. In this case, the osmotic pressure of clean water in hydrogel must be larger than that of the saline water. Li et al.[7,8] tested different hydrogels such as poly(sodium acrylamide) (PSA), poly(sodium acrylate)-co-poly(N-isopropylacrylamide), PAM and NIPAM, and PNIPAM, and found that PSA can draw most water. For such applications, no mobile ions exist in hydrogel. In a different forward osmosis approach, Yu et al.[50] directly immersed hydrogel made of poly(sodium acrylate-co-hydroxyethyl methacrylate) in saline water containing NaCl, and squeeze out water inside hydrogel with pressure. The recovered water contains less salt than the saline water. Thermodynamic analysis had been reported for such pressure-driven desalination cycle.[51,52] Rud et al.[53] simulated in detail the ion concentrations inside and outside hydrogel using a Monte Carlo method motivated by Yu et al.'s experiment. I will use the thermodynamics model presented above to compare the osmotic pressure difference between saline water and water in hydrogel.

The above experiments will be discussed under three categories: (1) Non-polyelectrolyte hydrogel such as PVA in equilibrium with pure water or humid air; (2) Non-polyelectrolyte hydrogel in equilibrium with salty water or hydrogel impregnated with salt in equilibrium with humid air; (3) Polyelectrolyte hydrogel in equilibrium with salty water or humid air.

**4. Non-electrolyte Hydrogel in Equilibrium with Pure Water or Humid Air**

Expressing the chemical potential in Eq. (20) at pressure p relative to ambient pressure $p_e$ as $\mu_w^*(T,p) = \mu_w^o(T,p_e) + v_w(p - p_e)$ and substituting the pressure difference using Eq. (13), we get the chemical potential of water inside the hydrogel as

$$\mu_w(T,p) = \mu_w^o(T,p_e) + RT[\ln(1 - f_2) + f_2 + \chi f_2^2] + RT\,K\left(f_2^{1/3} - f_2/2\right) \quad (23)$$

where $K = \frac{v_w n_e}{V_o} = \frac{v_1 v_e}{V_o}$. Recall $v_e$ is the effective number of crosslinking unit and $v_1$ the volume of a single solvent molecule, and $V_o$ the initial dry volume of the polymer. So K is the ratio of the effective volume of the crosslinking node to that of the dry volume, and it can be controlled by the crosslinker concentration in hydrogel synthesis. Equation (23) is identical to Flory's result except that it is clear that the last term is due to the pressure dependence of the chemical potential of pure water. Figure 1(a) plots the normalized excess chemical potential $[\mu_w(T,p) - \mu_w^o(T,p_e)]/(RT)$ as a function of the polymer volume fraction $f_2$ at different $\chi$ and K values. We see that the excess chemical potential from the ambient water, $\mu_w(T,p) - \mu_w^o(T,p_e)$ could be positive between [0,$f_{2,eq}$] under proper parameter combinations, which means that hydrogels are not stable and phase separation due to spinodal decomposition happens at $f_{2,eq}$, where $\mu(f_2 = 0) = \mu(f_2 = f_{2,eq})$, i.e., the polymer with some water ($f_{2,eq} = V_o/V$) is in equilibrium with pure solvent. From Eq. (23), at equilibrium, $\mu_w(T,p) = \mu_w^o(T,p_e)$, we have

$$[ln(1 - f_{2,eq}) + f_{2,eq} + \chi f_{2,eq}^2] + K\left(f_{2,eq}^{1/3} - f_{2,eq}/2\right) = 0 \qquad (24)$$

Figure 1(b) shows the equilibrium volume expansion of hydrogel as a function of K under different parameters. The smaller is K, i.e., the number of crosslinkers, the larger is the volume expansion due to the smaller tension of the polymers to counter the osmotic pressure of the solvent. We estimate that the PVA used in solar-driven interfacial evaporation[4] has K~0.03 and V/V$_o$~10. Reported value of $\chi$ for PVA is ~0.5,[54] depending on temperature and volume fraction. The calculated value is of the same order as the experiment.

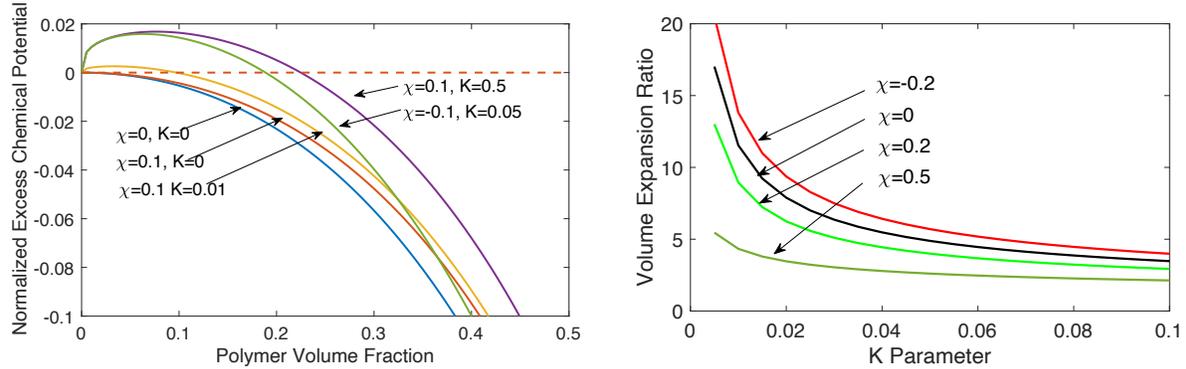

Figure 1 (a) Normalized excess chemical potential of water $[\mu_w(T,p) - \mu_w^o(T,p^o)]/(RT)$, under different parameters, (b) equilibrium volume expansion ratio as a function of parameter K under different $\chi$ values.

**Latent Heat of Evaporation.** Inside hydrogel, water is at different pressure p and hence its latent heat will be different from that of the solvent in the ambient pressure $p_e$. From the chemical potential of water, we derive that latent heat of evaporation is (see SI)

$$L = L_o + RT[ln(1 - f_{2,eq}) + f_{2,eq} + \chi f_{2,eq}^2] + RT^2 f_{2,eq}^2 \frac{\partial \chi}{\partial T} \qquad (25)$$

where $L_o = h_{w,v}^*(T,p_e) - h_w^*(T,p_e)$ is latent heat of evaporation of pure water at temperature T and $p_e$, and h the enthalpy with the subscript w for water and additional subscript v representing the vapor phase.

In Fig.2, we plot the latent heat as a function of $f_2$ for different $\chi$ values, assuming it is temperature independent. In reality, $f_{2,eq}$ is determined by $\chi$ and K as shown in Fig.1b. During differential scanning calorimetry (DSC) or solar-driven interfacial evaporation experiments,[3] hydrogel may not be in equilibrium and its chemical potential may be higher than that of the surroundings. We can use Eq. (25) to estimate the equivalent latent heat change as $f_2$ varies from small value to large value (when the sample is close to dried up). We see that when $f_2$ is large, the equivalent latent heat reduction change is appreciable. The latent heat reduction can be 1-$5\times10^5$ J/kg for polymer volume fraction between 90-99%. Such a level of latent heat change could be seen in DSC measurements. However, in solar-driven interfacial evaporation experiments using hydrogel,[3,4] the reported water evaporation rate can be ~3 times that of theoretical limit based on a nominal latent heat of 2.45 MJ, meaning a reduction of latent heat by 1.6 MJ, which is an order of magnitude larger than the pressure effect discussed here. The figure also shows that with increasing $\chi$ parameter, the latent reduction becomes smaller, because positive Flory-Huggins parameter means endothermic reaction when water binds to polymer, reducing the impact of higher pressure inside hydrogel. Hence, it does not seem that the pressure-change-caused evaporation latent heat reduction can explain the solar interfacial evaporation experimental results.

The higher enthalpic state of water inside the hydrogel due to higher pressure also means that when water diffuses into hydrogel, heat is absorbed despite that binding water with polymer is endothermic. If water enters the hydrogel from one surface and leaves from another surface, there is heat absorption at the interface where water enters and heat release at the interface where wate leaves the hydrogel. The absorbed and rejected heat is given by the same expression as Eq. (25). This is an effect similar to the Peltier effect in thermoelectric materials, although the later is due to the difference of electron's entropy flux,[55] while here it is due to the enthalpy difference of water molecules. In theory, a molecular Peltier cooling device can be built by allowing water to diffuse in from one side and diffuse out from the other through, for example, pressure or thermal driven flow. On a per molecule basis, $10^5$ J/kg heat absorbed is equivalent to 0.019 eV per molecule, or an effective Seebeck coefficient of 62.5 $\mu$V/K. This is comparable to typical electronic Seebeck coefficient. However, the value is not large enough to make such a cooling technology in a configuration similar to that of thermoelectric devices competitive because the small flow rate of the molecules, i.e., the equivalent low molecular conductivity, compared to that of electrons, although the intrinsic low thermal conductivity of the hydrogel is an advantage.[56] We had observed a small cooling effect when PVA hydrogels similar to that used in referenced solar-driven interfacial-evaporation experiment was first wetted with water, although we had not tested steady-state operation.

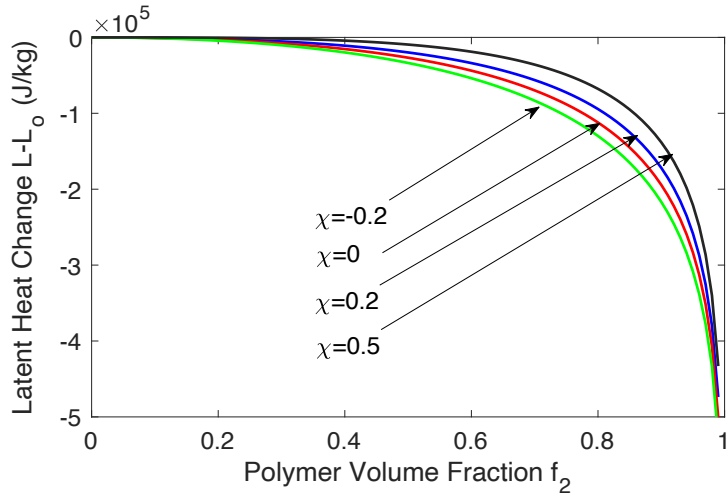

Figure 2 Latent heat reduction at T=300 K under different $\chi$ values as a function of the polymer volume faction.

**Freezing Point Depression and Boiling Point Elevation.** Next, we discuss the freezing point depression and the boiling point elevation, both were observed from DSC measurements on PVA hydrogels used in solar-driven interfacial evaporation experiment.[4,26] These observations can be explained based on the freezing point depression and boiling point elevation in solutions, typically called colligative properties for ideal cases when the ratios (melting point or boiling point) depend only on the concentration, and independent of the individual solute's chemical properties.[39,40] In fact, Huggins[34] had used the freezing point depression in developing his theory on polymer solutions. The standard approach in deriving the freezing point depression and boiling point elevation is to set the differentials of the chemical potential, i.e, the change in chemical potential of water inside the gel equaling to that of the solid (for freezing point) or pure vapor (for boiling point), at which the activity can be taken as one. However, in typical DSC experiments, the hydrogel with water inside is taken out of the water bath. If one waits long enough, hydrogel will establish new equilibrium with the ambient through volume change. The mechanical equilibrium with the ambient at pressure $p_o$ is ensured by Eq. (24). The chemical equilibrium will be established by setting water chemical potential equaling the chemical potential of water in the ambient gas (air or inert gas used in DSC measurements). We will discuss more of such an equilibrium in water adsorption section. For now, we will take $f_2$ as a variable. At least in the DSC evaporation experiment, hydrogel is continuously evaporating and is not in chemical equilibrium with the ambient.

Following the standard approach as discussed above,[38,39,40] we obtain the following expression for the freezing point ratio (see SI)

$$\frac{T_{fz,0}}{T_{fz}} \approx 1 - \frac{RT_{fz,0}}{L_m}[ln(1-f_2) + f_2 + \chi f_2^2] \tag{26}$$

and the boiling point (see SI)

$$\frac{T_{bp,o}}{T_{bp}} - 1 = \frac{RT_{bpo}}{L_o}\left\{ln(1-f_2) + f_2 + \chi f_2^2 + K\left(f_2^{1/3} - \frac{f_2}{2}\right)\right\} \tag{27}$$

where $T_{fz,0}$ is the metling point and $T_{bp,o}$ the boiling point of pure water when $f_2$=0. We take a value of $L_m$=334 kJ/kg and Lo=2.45 MJ/kg, and plot the freezing point $T_{fz}$ and boiling point $T_{bp}$ as a function of $f_2$ as shown in Fig.3(a) and 3(b) for different values of $\chi$. We see indeed that the freeze point is suppressed. We estimate that $f_2$~0.7 when freezing depression ~20 °C was observed,[26] which is comparable to value predicted. Figure 3b shows the boiling point elevation. The plot starts from certain value of $f_2$, which is the equilibrium volume for $f_{2,eq}$ that is the solution of Eq. (24). DSC experiment typically measure a broaden evaporation temperature range which can exceed the boiling point of pure water by ~60 °C,[3] consistent with what we see from Fig. 3b.

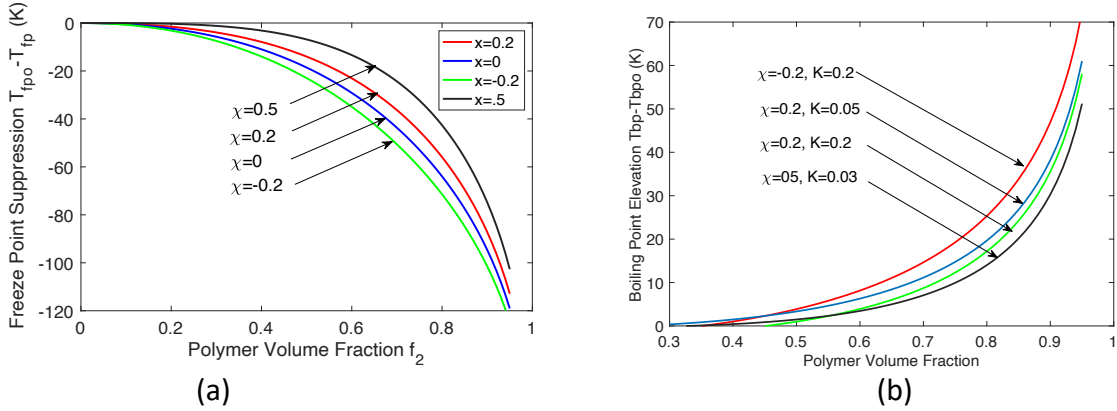

(a)    (b)

Figure 3. Calculated (a) freeze point depression from 0 °C and (b) boiling point elevation from 100 °C.

In passing, we make the comment that if boiling happens inside the hydrogel, the vapor pressure will be even higher than p. One can combine the Laplace-Young equation for the vapor pressure and obtain corresponding boiling point inside hydrogel following a similar approach.

**Trapping Water from Air.** Let us now consider if a hydrogel can trap water when it is placed in air. This is of interests to atmospheric water harvesting as well as to maintain functions of a hydrogel when it is exposed to air.[9,27] The chemical potential of water vapor in air is related to water vapor's partial pressure $p_v$ relative to its saturation pressure $p_s$ at temperature T (assuming air is an ideal mixture),

$$\mu_{w,v}(T, p_v) = \mu^*_{w,v}(T, p_s) + RT ln\left(\frac{p_v}{p_s}\right) \tag{28}$$

We equate the above chemical potential to water chemical potential inside hydrogel, i.e., Eq. (23), to get

$$\left[ln(1 - f_{2,eq}) + f_{2,eq} + \chi f_{2,eq}^2\right] + K\left(f_{2,eq}^{1/3} - \frac{f_{2,eq}}{2}\right) = ln\left(\frac{RH}{100}\right) \tag{29}$$

where RH is the relative humidity. When hydrogel is emersed in water, the left hand side equals zero, which is Eq.(24). However, when a piece of hydrogel is taken out of water, it will arrive at a new equilibrium determined by Eq. (29). Since the right hand side is significantly more negative at low RH, the $f_{2,eq}$ value will increase significantly, i.e, the hydrogel will shrink. In Fig.4, we show the water volume fraction $(1-f_2)$ as a function of the RH for different parameters. The value at RH=100 is also the equilibrium water content when hydrogel is immersed in water since the chemical potential of water in saturated vapor equals that of the pure liquid water. For a given hydrogel, $\chi$ is fixed. One can reduce K to trap more water, i.e., use less cross-linkers. At low humidity, however, there is only very little water left in the hydrogel. In completely dry environment, pure nonelectrolyte hydrogel cannot retain water. These are consistent with the challenge in retaining water in hydrogel.

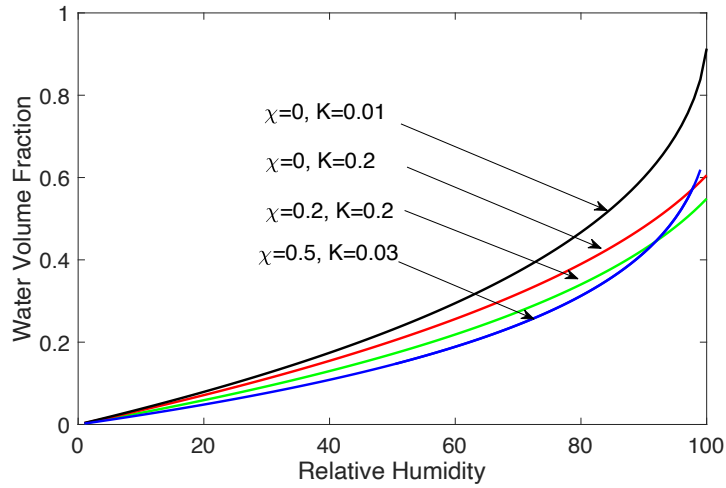

Figure 4. Water content (volume fraction) in hydrogel as a function of relative humidity for different parameter combinations.

## 5. Nonpolyelectrolyte Hydrogels with Salt In Equilibrium with Water or Humid Air

This topic is of interests for several applications we mentioned before. One of them is in solar-driven interfacial evaporation using hydrogels. It was found that there is no salt precipitation inside the hydrogel,[3,4] an antifouling property. The other is that by incorporating salt inside hydrogel, it is found that hydrogel does not dry out.[27]

**Salt Content Inside Hydrogel in Equilibrium with Saline Water Outside.** Let us consider first for simplicity a symmetric monovalent salt such as NaCl salt is added to water outside the hydrogel. To find out the salt concentration inside the hydrogel at equilibrium, we start with Eq. (21) to write done the chemical potential of Na$^+$ and Cl$^-$ ions inside the hydrogel as

$$\mu_{Na+} = \mu_{Na+}^o + v_{Na+}(p - p_e) + F\varphi + RT[\ln(1 - f_2) + f_2 + \chi f_2^2 + \ln(\gamma_{Na+} x_{Na+})] \quad (30)$$

$$\mu_{Cl-} = \mu^o_{Cl-} + v_{Cl-}(p - p_e) - F\varphi + RT[\ln(1 - f_2) + f_2 + \chi f_2^2 + \ln(\gamma_{Cl-}x_{Cl-})] \quad (31)$$

where $\varphi$ is the potential difference (inside minus outside of hydrogel), $v_{Na+}$ and $v_{Cl-}$ are molar volumes of Na$^+$ and Cl$^-$ ions, and $x_{Na+}$ and $x_{Cl-}$ are mole fraction of Na$^+$ and Cl$^-$ ions inside the hydrogel, respectively. Due to the inclusion of electrostatic potential, the chemical potential as given in Eqs.(30) and (31) is actually the electrochemical potential. Each type of ions is in chemical equilibrium with the same type of ions outside, and their chemical potentials equal each other. This is the basic strategy used in Donnan equilibrium to obtain the equilibrium concentrations. Typical Donnan equilibrium treatment as Eq. (5), however, neglects the pressure term in the chemical potential.[37] To find this pressure, we also need to employ the chemical potential balance of water inside and outside. The final equations to solve are (see SI)

$$K_{NaCl}\left[f_2^{1/3} - \frac{f_2}{2}\right] + 2[\ln(1 - f_2) + f_2 + \chi f_2^2] + \ln(\gamma_{NaCl}x_{NaCl})^2 = \ln[(\gamma_{NaCl}x_{NaCl})^2]_e \quad (32)$$

$$K\left[f_2^{1/3} - \frac{f_2}{2}\right] + [\ln(1 - f_2) + f_2 + \chi f_2^2 + \ln(\gamma_w x_w)] = \ln(\gamma_w x_w)_e \quad (33)$$

where $K_{NaCl} = \frac{n_e v_{NaCl}}{V_o}$ with $v_{NaCl} = v_{Na+} + v_{Cl-}$, and subscript e represents external solution. Although we did assume that volume of ions equal to that of water molecule in the entropy of mixing expression in Eq. (21), we will take here that K$_{NaCl}$ can differ from K for water as it is the case based on the molar volume measurements of salts.[57,58,59] This is acceptable since for the entropy of mixing, water usually dominates over ions, while the ion volume directly enters the pressure term here.

In Fig.5(a), we show the equilibrium polymer volume fraction and Fig.5(b) the difference of the ion mole fraction between the external and the internal of the hydrogel, as a function of external ion mole fraction, varying different parameters but taking activity coefficient γ=1 for all species.

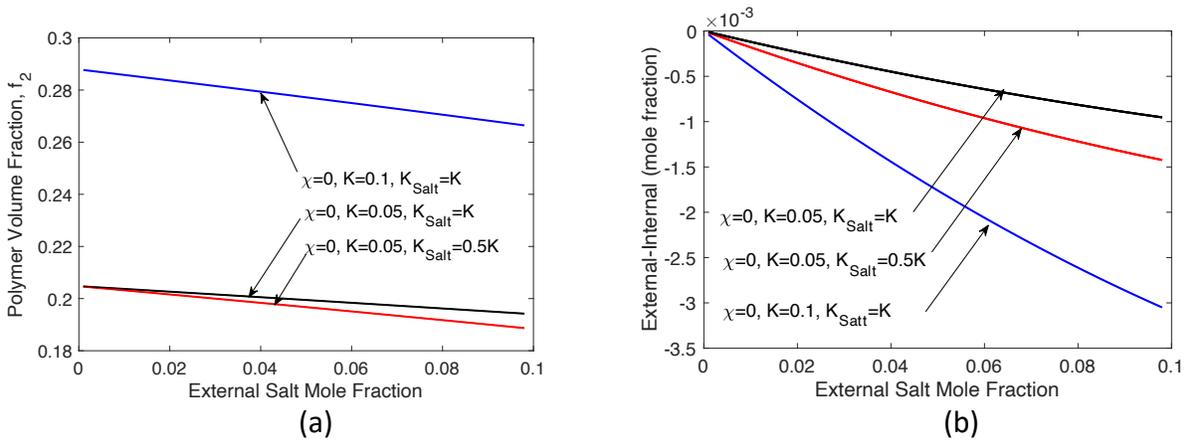

(a) (b)

Figure 5 Parametric studies of hydrogel in monovalent symmetric salt solution (a) hydrogel volume fraction and (b) external to internal salt mole fraction difference as a function of the external salt mole fraction. All calculations assume the activity coefficients equaling one.

We see that as the external ion mole fraction increases, $f_2$ decreases, i.e., the hydrogel expands (as long as $K_{NaCl}<2K_{H2O}$, which is usually the case for ions). The external salt content is less than internal salt content [Fig.5(b)]. One can prove this rigorously by multiplying Eq. (33) by 2 and subtracting from it Eq. (32). In terms of a physical picture, under the current model, we see from Eqs. (19) and (32) that every NaCl-type of salt enters hydrogel is equivalent to two water molecules albeit with a different volume. The smaller is the molar volume of the salt, the less is the pressure term contributions. More salt ions exist inside to counter the loss of pressure term contribution. More salts inside also create a higher osmotic pressure, which explains why the hydrogel expands.

However, real salts differ from each other. The difference is reflected in their activity coefficients, which vary with salt concentration.[41] In addition, the molar volume of salts also changes with concentration.[59] For water, rather than the activity coefficient, the osmotic coefficient defined as the ratio of $\phi = ln(a_w)/ln(x_w)$ is often given. We took the measured[60] activity coefficients of NaCl and LiCl salts in water and the corresponding osmotic coefficient (Fig.6(a)], and recalculated the same quantities, as shown in Figs.6(b) and 6(c). The partial molal volumes of NaCl and LiCl are 16.8 and 16.6 cm³/1 kg and hence we took $K_{salt}/K$=0.3.[57] We neglect the concentration dependence of the partial molar volume in these calculations. We see that LiCl has higher activity coefficient and osmotic coefficient, which leads to larger volume expansion and concentration difference. For LiCl, internal salt concentration is higher for most concentrations, while for NaCl, the external salt concentration is higher for most concentrations.

Also, in this case, a membrane potential might exist, which can be calculated from the following equation (SI)

$$(K_{Na+} - K_{Cl-})\left[f_2^{1/3} - \frac{f_2}{2}\right] + \frac{2F\varphi}{RT} = ln\left[\frac{ln(\gamma_{Cl-}x_{Cl-})(\gamma_{Na+}x_{Na+})_e}{(\gamma_{Na+}x_{Na+})(\gamma_{Cl-}x_{Cl-})_e}\right] \tag{34}$$

where $K_{Na+} = \frac{n_e v_{Na+}}{V_o}$ and $K_{Cl-} = \frac{n_e v_{Cl-}}{V_o}$. Typically, the molar volume of anions are larger than cations. and it is unlikely that first term on the left-hand side will exactly cancel the right hand side term, implying that a potential difference could develop between inside and outside hydrogel.[37] We plotted in Fig.6(d) the potential difference between inside and outside the hydrogel. This potential difference should not exist under standard treatment of the Donnan potential since there are no fixed charge inside the gel. We predict the existence of a potential in this case because the pressure difference between inside and outside the hydrogel. Although we have not found any publications reporting such a potential difference, I learnt from private communication that a potential indeed exists between nonelectrolyte hydrogel emmersed in salt.[49]

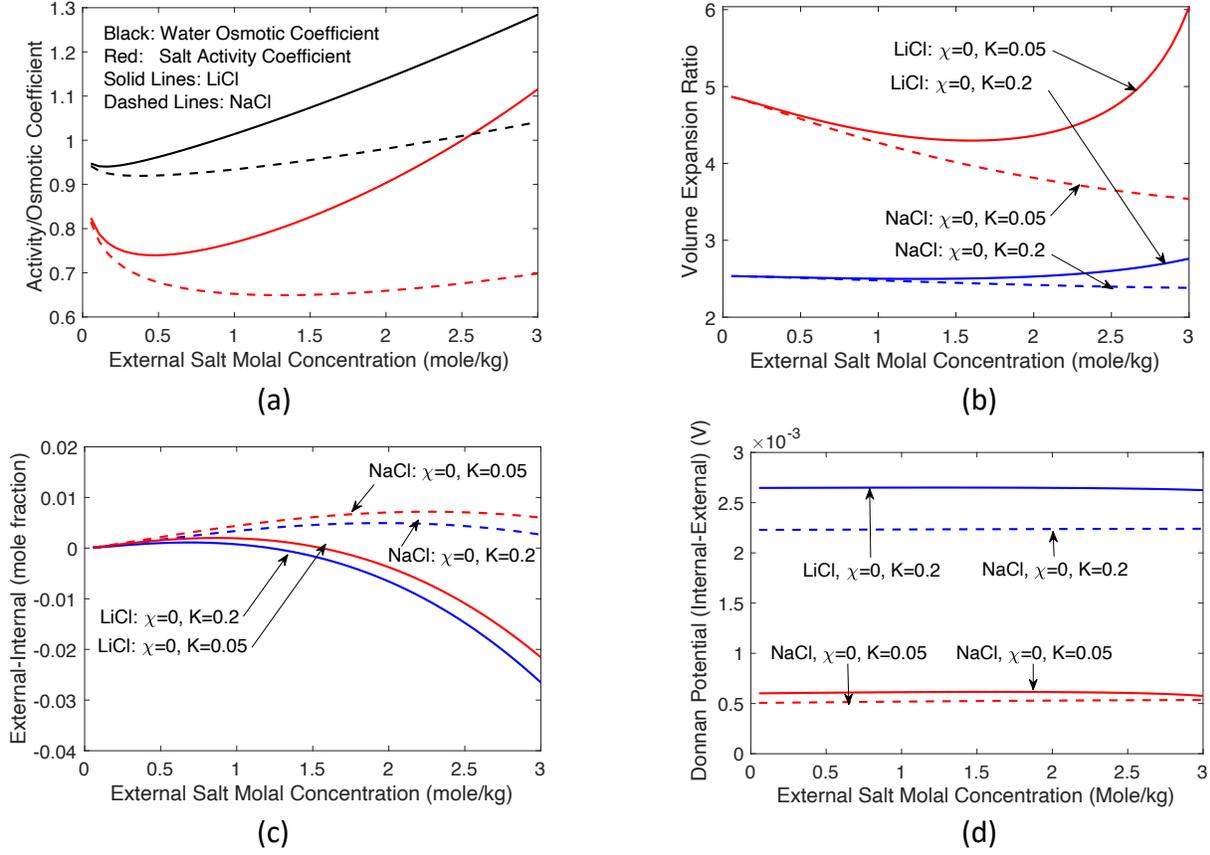

Figure 6. Impact of salt characteristics on volume expansion and salt concentration (a) activity and osmotic coefficients of NaCl and LiCl (data from Ref.60), (b) volume expansion ratio, (c) salt concentration difference, and (d) Donnan potential as a function of the external salt molal concentration. Standard Donnan potential theory predicts a zero potential because there is no fixed charge.

**Trapping Water from Air.** If a hydrogel impregnated with a salt is in equilibrium with air, the water chemical potential balance leads to

$$K\left[f_2^{1/3} - \frac{f_2}{2}\right] + [ln(1-f_2) + f_2 + \chi f_2^2 + ln\gamma_w x_w] = ln\left(\frac{RH}{100}\right) \qquad (35)$$

If the salt volume is $V_s$ and the dry polymer volume is $V_o$, we can relate the salt mole fraction to the polymer volume fraction in expanded hydrogel as $x_s \approx \frac{V_s}{V_o(1/f_2-1)}$, where the approximate sign is due to our assumption of equal volume between water molecule and ions in the entropy of mixing formula. The water mole fraction is $x_w = 1 - 2x_s$, where the factor of 2 accounts for two different ions. We solved Eq. (35) using the water osmotic coefficients for NaCl and LiCl and results are shown in Figs.7(a) and 7(b). Figure 7(a) shows the amount of water as a function of the relative humidity. Compared to Fig.4 for hydrogel not containing any salt, hydrogel with salt can absorb more water even at low humidity. Figure 7(b) shows the water mole fraction. The solubility of salt determines the minimum humidity that the salt will be dissolved in water. LiCl can maintain more water due to its (1) larger activity coefficient and (2) higher solubility. Bai et al.[27] tested the water-retention ability of PAM hydrogel impregnated with LiCl and NaCl and found that NaCl precipitates out while LiCl impregnation can maintain hydrogel in wet state. Our results here, together with the fact shown in Fig. 6(c) that there are more NaCl outside than inside in a solution, are consistent with the trends of experimental observations.

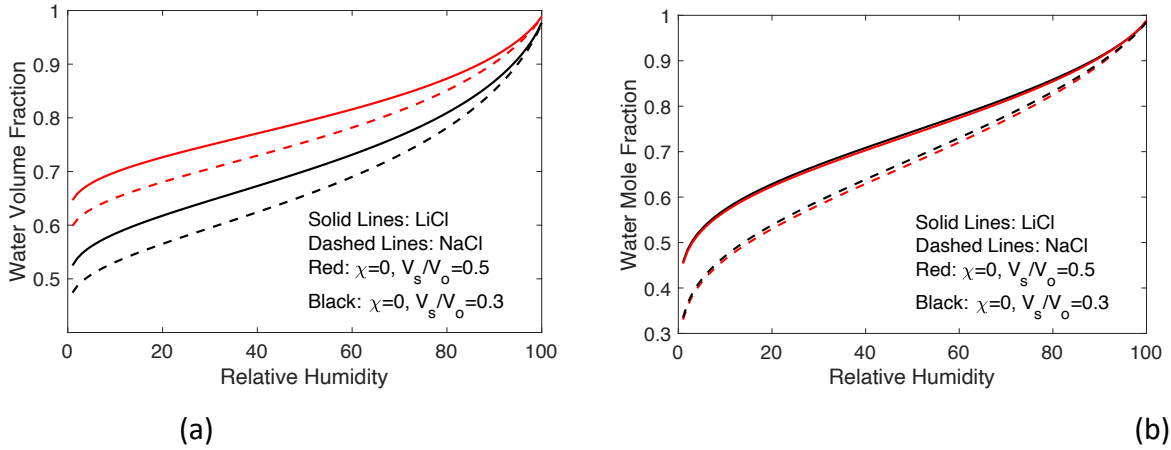

(a)                                                          (b)

Figure 7. Comparison of atmospheric water extraction, i.e. hydrogel water retention performance for hydrogel impregnated with LiCl and NaCl (a) water volume fraction and (b) water mole fraction as a function of humidity. The salt solubility in water set the lower limit of RH. Clearly, LiCl has better performance than NaCl due to its larger activity coefficients.

**Solubility of Salts inside Hydrogel.** The above analysis shows whether there is more salt inside or outside depends on the type of salt: more NaCl outside while more LiCl inside. While this might explain the antifouling properties observed in solar-driven hydrogel experiment because NaCl is the main salt in saline water, there are also two other possible mechanisms that can affect the salt fouling: (1) mass transfer,[61] and (2) salt might have higher solubility inside the hydrogel. The analysis of solubility is identical to that of the freezing point depression.[38] One can think for example that a salt such as NaCl is at equilibrium between pure solid and pure liquid form at its melting temperature $T_m$=801 °C. The fact that at room temperature, there is equilibrium between solid NaCl in a solution can be considered as the melting point depressed from $T_m$=801 °C to room temperature. Following the established method, we obtain the following relation for the solubility $x_{NaCl}$ (see SI):

$$-\frac{L_m}{2RT}\left(1 - \frac{T}{T_m}\right) \approx [\ln(1-f_2) + f_2 + \chi f_2^2] + \ln(\gamma x_{NaCl}) \qquad (36)$$

This equation is similar to freezing point depression Eq. (26), except a factor of 2 due to the fact that Na⁺ and Cl⁺ ions are different and are treated separately. Comparing the above expression against the case $f_2$=0, i.e., without the square bracket term on the right hand of the above equation, we see that hydrogel increases solubility of salt since the square bracket term is negative. This increase at first might seem to be a little surprising. On the other hand, we can understand it as follows. As temperature drops, the liquid phase Gibbs free energy will increase, this is why there is no liquid of NaCl at lower than the melting temperature. In a solution, the partial Gibbs free energy of the salt is reduced by the mixing term, which makes it possible for the solute to exist. With polymer, the solute entropy further increases, so more solute can exist, i.e., the solubility is higher.

Neglecting the activity coefficient dependence on concentration, the solubility ratio is

$$\frac{x_{NaCl}(in\ hydrogel)}{x_{NaCl}(without\ hydrogel)} = \frac{1}{(1-f_2)exp[f_2+\chi f_2^2]} \tag{37}$$

Figure 8 shows the solubility ration as a function of polymer volume $f_2$. The increased solubility of salt inside hydrogel could further contribute to why hydrogel has antifouling properties, together with mass diffusion arising from the concentration difference.[61]

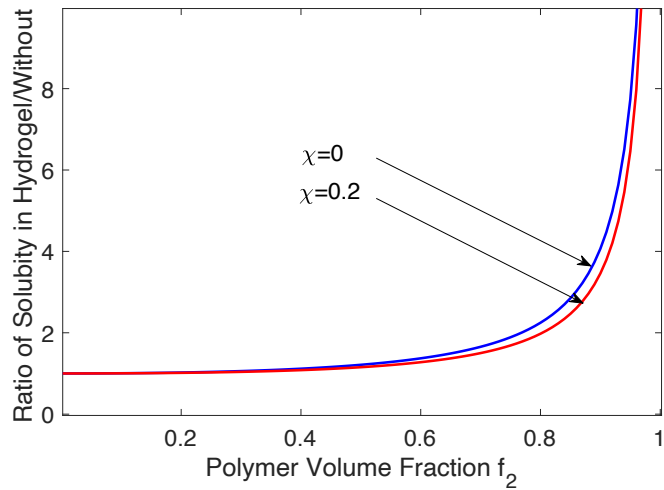

Figure 8  Ratio of solubility inside hydrogel vs. without hydrogel as a function of the polymer volume fraction.

6. **Polyelectrolyte Hydrogels in Equilibrium With Water Containing Salt or Humid Air**

Let us consider now a polyelectrolyte hydrogel, for example, poly sodium acrylate (PSA). Polyelectrolyte hydrogels had been used to extract water from atmosphere[9] and retain water in hydrogel,[27] and as a drew agent in forward osmosis for desalination applications. The forward osmosis application employs semipermeable membranes to separate ions. Here we first consider the case with no membranes, which will be of interests to solar evaporation and atmospheric water harvesting applications we discussed before.

We assume that the equilibrium constant for the monomer dissociation

$$CHCOONa \rightarrow CHCOO^- + Na^+ \tag{38}$$

can be applied to polymer and hydrogel too. The activity-based equilibrium constant can be written as

$$K_A = \frac{a_{AA^-} \times a_{Na^+}}{a_{SA}} = \frac{n_{AA^-} \times n_{Na^+}}{(n_w + n_{AA,t} + n_{Na^+} + n_{Cl^-})(n_{AA,t} - n_{AA^-})} \quad (39)$$

where $n_{AA^-}$ is the number of ionized monomers and $n_{AA,t}$ is the total number of monomers in the hydrogel. In the second equality, we assumed the activity coefficients equaling one and accounted both the solvent molecules and ions in the calculation of the mole fraction. Since the pressure inside hydrogel is higher, the equilibrium constant depends on pressure[38,39]

$$K_A(T, p) = K_A(T, p_o) exp\left[\frac{(v_{AA^-} + v_{Na^+} - v_{PAA})(p - p_o)}{RT}\right] \quad (40)$$

For the following, we neglect the pressure correction in the last term of Eq. (40), which is reasonable if the molar volumes of dissociated ions do not differ from that of the nondissociated monomers. From Eq. (39), we have

$$\frac{n_{AA^-}}{n_{AA,t}} = \frac{K_A(n_w + n_{AA,t} + n_{Na^+} + n_{Cl^-})}{K_A(n_w + n_{AA,t} + n_{Na^+} + n_{Cl^-}) + n_{Na^+}} \quad (41)$$

The chemical equilibrium in hydrogels are more complicatet than modeled here because the ionized nAA- can also be neutralized by H$^+$, which will set a limit to the dissociation of weak polyelectrolyte. We refer readers to more rigorous treatment in literature.[21]

**Polyelectrolyte Hydrogel in Equilibrium with Brine.** Consider now the hydrogel is in equilibrium with outside water containing NaCl salt. Inside the hydrogel, we have fixed AA$^-$ ions and mobile Na$^+$ and Cl$^-$ ions. Charge neutrality requires

$$n_{Na^+} = n_{Cl^-} + n_{AA^-} \quad (42)$$

We assume that the equivalent volume of a monomer to water is y, then $y n_{AA,t} = Z n_2$. Dividing the above equation by $n_{Na^+} + n_{Cl^-} + n_w$ and using Eq. (41), we arrive at,

$$x_{Na^+} = x_{Cl^-} + \frac{K_A\{1 + f_2/[y(1-f_2)]\} f_2/[y(1-f_2)]}{K_A[1 + f_2/[y(1-f_2)]] + x_{Na^+}} \quad (43)$$

To find f$_2$ and the salt concentration inside hydrogel, we balance the chemical potential for each mobile species as in the previous section (see SI). Figure 9 plots solutions for different equilibrium constant values while holding other parameters constant.

Compared to the case of non-polyelectrolyte hydrogels (Fig.6) immersed in salty water, we see that with immobile ionized polymer chains inside hydrogel, the salt is repelled from the hydrogel for most cases. When the external salt is less, i.e., the gel expands more. A peak appears in Fig.9(a) for the difference of external to internal salt concentration when the equilibrium constant is small, i.e., the dissociated mobile ions' concentration is low. This peak can be understood as follows: with more ionization of the polyelectrolyte and hence more fixed charge inside ($K_A$ large), salt is expelled. With less ionization of polyelectrolyte, Figure 5b shows there are more salt inside. These two different trends lead to the existence of the peak. The Donnan potential [Fig.9(c)] is one order of magnitude larger than non-electrolyte hydrogel immersed in salt solution [Fig.6(d)].

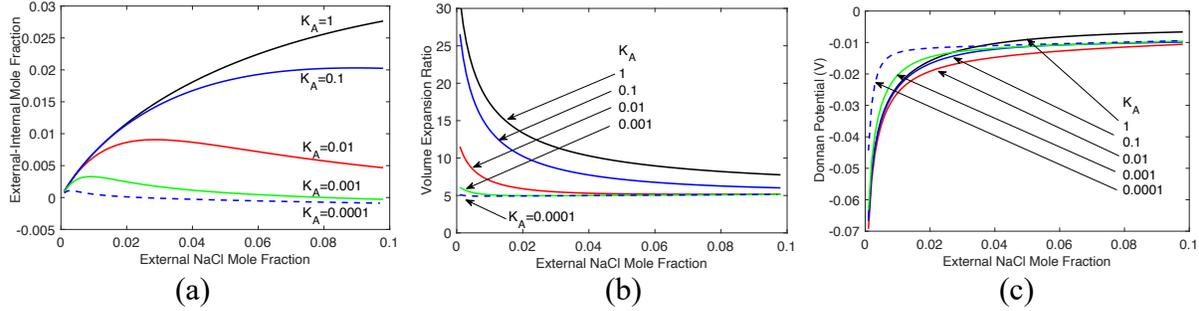

(a) (b) (c)

Figure 9 Polyelectrolyte hydrogel immersed in monovalent symmetric salty water (a) salt concentration difference, (b) polymer volume fraction, and (c) Donnan potential as a function of the external salt concentration ($\chi=0$, $K=0.05$, $K_{salt}=K$, $y=2$, $\gamma_i=1$).

**Polyelectrolyte Hydrogel for Water Trapping from Air.** In terms of the capability of electrolyte hydrogel trapping water in air, Eq.(35) is still applicable. For polyelectrolyte hydrogel, the dissociated ions ($Na^+$ using the example we consider here) mix with water, which increases the entropy of the solution inside hydrogel. Water mole fraction can be obtained from solving Eq. (43) with $x_{Cl^-}=0$, which leads to the ionized sodium mole fraction and consequently the water mole fraction as,

$$x_w = 1 - \frac{K_{PSA}}{2}\left[1 + \frac{f_2}{y(1-f_2)}\right]\left\{\left[1 + \frac{4f_2}{K[f_2+y(1-f_2)]}\right]^{0.5} - 1\right\} \quad (44)$$

Figure 10 shows the water volume at different parameter values obtained from solving Eqs. (35) and (44), focusing on changing the equilibrium constant. We can see that the larger the equilibrium constant, i.e, the more polyelectrolyte that can be ionized, the more water the hydrogel can contain. Compared to impregnate salt into non-electrolyte hydrogels (Figs.6 and 7), the latter are limited by the salt solubility. Using electrolyte hydrogels will be more effective in retaining water or extracting water from air. Strong polyelectrolyte with large equilibrium constant is desired.

**Forward Osmosis.** Forward osmosis using hydrogels or uncrosslinked polyelectrolytes as the draw solution.[7,8] The osmotic pressure of hydrogel in fresh water at equilibrium inside the hydrogel should be larger than that of the salt water separated by a membrane. For polyelectrolyte hydrogel, its equilibrium volume can be obtained from solving

$$K\left[f_2^{1/3} - \frac{f_2}{2}\right] + [\ln(1-f_2) + f_2 + \chi f_2^2 + \ln(\gamma_w x_w)] = 0 \quad (45)$$

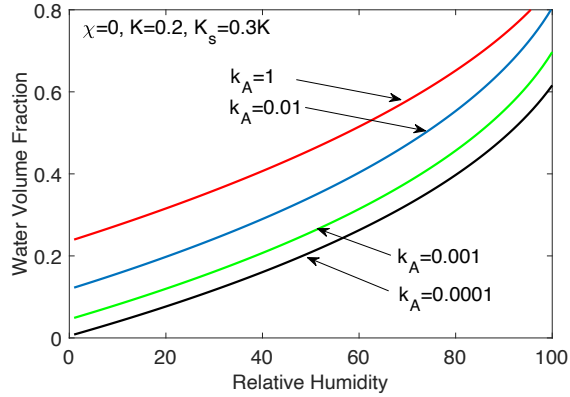

Figure 10. Comparison of water content vs. humidity at different parameters ($y=2$, $\gamma_i=1$).

where water mole fraction is again given by Eq. (44). Solving Eqs. (44) and (45), we obtain the equilibrium volume $f_{2,eq}$. The condition for forward osmosis to happen is then

$$K\left[f_{2,eq}^{1/3} - \frac{f_{2,eq}}{2}\right] > 2x_{salt} \tag{46}$$

where $x_{salt}$ is the brine solution salt concentration and we assumed monovalent symmetric salt so that the brine's osmotic pressure is approximately $2RTx_{salt}/v_w$.

Figure 11(a) shows the equilibrium hydrogel volume expansion ratio at different dissociation equilibrium constant values as a function of K, which can be controlled by crosslinking. Compared to non-electrolyte hydrogels, electrolyte hydrogels have much large swelling ratio, a well-known fact. Figure 11(b) shows the osmotic pressure. Sea water osmotic pressure is 25-35 bar, as marked in shaded region. We can see that electrolyte hydrogel osmotic pressure can be much higher, suggesting that electrolyte hydrogel can serve as a good draw medium for forward osmosis, which is consistent with the trend observed in experiments.[7,8] The higher is the $K_A$ value, the larger is the equivalibrium volume, i.e., more clean water can be obtained. For regeneration of hydrogel, temperature sensitive hydrogel such as Poly(N-isopropylacrylamide) (PNIPAM) had been used,[7,8] which can be modeled by including the temperature dependence of $\chi$ parameter.[62]

**Multivalent Ions.** Although the examples we gave above took monovalent symmetric salts, extension to unsymmetric and polyvalence ions are straightforward, which have been well-explained in textbooks.[38,39] For $MgCl_2$ salt, for example, for each mole of $MgCl_2$ salt, there are two moles of Cl- ions and one mole of Mg ions ($z_i=1$ for $Mg^+$ and 2 for Cl- in Eq.(17). Similar operation as arriving at Eq. (33) will lead to a term like $ln(\gamma_{MgCl2}2x_{MgCl2})^3$, in which the factor of 2 is due to the fact that the mole fraction of Cl- ions are twice of $x_{MgCl2}$. The factor 2 in front of the mixing term in Eq. (32) will be replaced by 3.

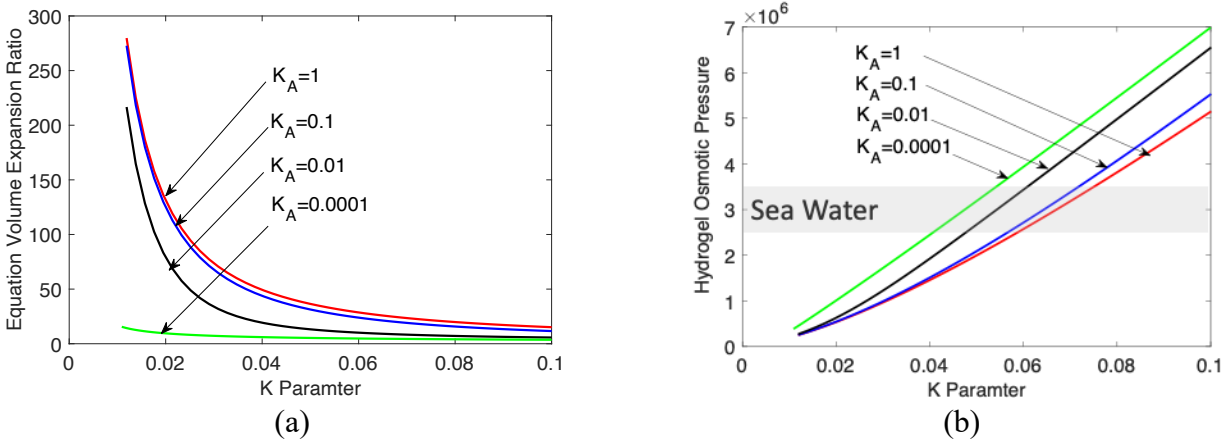

Figure 11. (a) Equilibrium volume expansion ratio of polyelectrolyte hydrogel at a function of K parameter at different values of the equilibrium constant $K_{PSA}$, and (b) osmotic pressure of the hydrogel ($\chi=0$, $y=2$).

## 7. Summary

To summarize, we have reformulated Flory's thermodynamic model for hydrogels based on the minimization of Helmholtz free energy of the combined hydrogel and its ambient. Our reformulation emphasizes the volume and molecular numbers as independent natural thermodynamic variables, which clarifies the roles of thermodynamic pressure. This formulation led to clear picture of the interplay between the osmotic pressure and the tension in polymer, and allows one to express the chemical potentials of the solvents and ions inside the hydrogel unambiguously. Relying on the chemical potential expressions, we examine water and ions in hydrogels related to applications in solar-driven interfacial evaporation, atmospheric water harvesting and water retention ability, and desalination. The chemical potentials of the solvent molecules and dissolved ions differ from that of their values in pure states, which lead to changes of the latent heat, boiling and freezing points, and solubility. The balances of each mobile species inside and external to hydrogels determine the osmotic pressure, Donnan potential, the water and salt contents inside the hydrogel. Based on the trends predicted by the model, we can draw the following conclusions.

Hydrogels have reduced latent heat of evaporation, due to increased pressure of water inside hydrogel. However, the reduction in the latent heat is not large enough to explain experimental observations. The melting point suppression and boiling point elevation phenomena as observed in DSC experiments can be explained similar to colligative properties observed in solutions, and they are caused by increased mixing entropy of water with polymer inside hydrogel.

When nonelectrolyte hydrogel is placed in salty water, salt contents inside hydrogel could be higher or lower than outside, depending on salt's activity coefficient and concentration. Salts generally have higher solubility inside hydrogel. Regular nonelectrolyte hydrogel can get completely dry in dry air. By impregnating salt, however, hydrogel can retain more water. Polyelectrolyte hydrogels have better ability to absorb water from moist air and retain water and such abilities increase the larger the ionization equilibrium constant. For electrolyte hydrogels immersed in salty water, the salt concentration inside is typically lower than outside.

We should point out that although the thermodynamic approach we presented is general, the specific models we used for the Helmholtz free energy are simplified. We rely mainly on Flory's theories on the mixing entropy between the solvent and polymer molecules, and polymer's configurational entropy changes for the Helmholtz free energy. We did not consider the interaction between mobile ions and the fixed ions on the polymer, for which large amount of work has been done and different models exist. We only included polymer's configurational entropy changes, excluding the intermolecular force interactions along the chain which could be important in highly stretched polymers. We did not treat the dissociation of polyelectrolyte rigorously. All these details could be improved and indeed many studies had already been done as we cited along the way. The thermodynamic treatment we present however is general, and improvements in these details can be incorporated to make better predictions. However, the classical thermodynamic treatment is by nature a mean field theory.

**Acknowledgments:**
I would like to thank Dr. Xueyu Li for her attempts in measuring Donnan potential in PVA hydrogel with salty environment and indeed observation of potential difference and Dr. Honglei Guo for sharing his PhD thesis on this topic. The author would like to thank Drs. Shaoting Lin, Jungwoo Shin, Yaodong Tu, and Mr. Simo Pajovic for helpful discussion, and for MIT to its support.

**Conflicts of Interests:**
The author declares no conflict of interests.

Supplementary Information

Thermodynamics of Hydrogel for Applications to
Atmospheric Water Harvesting, Evaporation, and Desalination

Gang Chen

Department of Mechanical Engineering
Massachusetts Institute of Technology

**Flory Theory.** Flory[12] developed a lattice model and arrived at expression for the entropy of mixing between a solvent and a polymer that is different from Gibbs' entropy of mixing expression. For a solution of $n_1$ molecules of the solvent and $n_2$ molecules of the polymer, the increase in entropy due to mixing is

$$\Delta S = -k_B(n_1 \ln f_1 + n_2 \ln f_2) \tag{S1}$$

where $\Delta$ means final minus initial values and $k_B$ is the Boltzmann constant, and f the volume fraction of the solvent and the polymer, which can be expressed as

$$f_1 = \frac{n_1}{n_1 + Z n_2}, \qquad f_2 = \frac{Z n_2}{n_1 + Z n_2} \tag{S2}$$

with Z denoting the volume ratio of the polymer molecule to the solvent molecule. Equation (S1) can also be derived from the free-volume concept.[13] Its main difference from the classical Gibbs entropy of mixing expression is that the volume fraction is used instead of the mole fraction. Equation (S1) degenerates into the Gibbs expression if $Z=1$.

In addition to the entropy of mixing, the mixing of the solvent with the polymer also causes a change in the enthalpy, which can be expressed as

$$\Delta H = k_B T \chi n_1 f_2 \tag{S3}$$

where $\chi$ is a parameter depending on the solvent and the polymer. $\chi > 0$ means the mixing is endothermic and $\chi < 0$ exothermic. Most polymer-solvent mixing has $\chi > 0$. We note here that although Eq.(S3) is called the enthalpy change, no pressure effect was taken into account in the derivation of the above expression. Inside hydrogels, the pressure is usually much higher than the ambient pressure. This pressure effect is not included in Eq. (S3).

For a polymer, Flory's[10] analysis of experimental data showed that the elasticity arises from the polymer configurational entropy change during stretching, for which Flory and Rehner[14] derived the following expression

$$\Delta S = -\frac{k_B \nu_e}{2}\left[a_x^2 + a_y^2 + a_z^2 - 3 - \ln(a_x a_y a_z)\right] \tag{S4}$$

where $a_x$ ($y$&$z$) is the stretching ratio in the direction represented by the subscript and $v_e$ is the effective number of crosslinked units that excludes the two free ends of a polymer chain. $v_e$ is related to the actual cross-linking units $v$ and the number of molecules $n_2$ through

$$v_e = v - 2n_2 = v(1 - 2n_2/v) = v(1 - 2M_c/M) \tag{S5}$$

where $M_c$ and $M$ are the molecular weight of a crosslinked segment and that of the polymer molecular before crosslinking, respectively. Combining Eqs.(S1), (S3), and (S4), the total change in the Gibbs free energy for isotropic swelling is Eq. (1).

**Helmholtz Free Energy of Combined System:** From Eq.(6), we can express the derivative of the Helmholtz free energy of the hydrogel subsystem as

$$dF = \left[\left(\frac{\partial(\Delta F_{mix})}{\partial V}\right)_{T,\varphi,n_j} + \left(\frac{\partial(\Delta F_{pl})}{\partial V}\right)_{T,\varphi,n_j} - p\right]dV$$

$$+ \left[\left(\frac{\partial(\Delta F_{mix})}{\partial T}\right)_{V,\varphi,n_i} + \left(\frac{\partial(\Delta F_{pl})}{\partial T}\right)_{T,\varphi,n_i} - S_c\right]dT$$

$$+ \mu_{sl}dn_{sl} + \mu_{pl}dn_{pl} + \sum_i \mu_i dn_i + \sum_i ez_i n_i d\varphi + ez_{pl}n_{pl}d\varphi \tag{S6}$$

Similarly, the Helmholtz free energy of the external solution can be written as

$$dF_e = \left[\left(\frac{\partial(\Delta F_{mix})}{\partial V}\right)_{T,\varphi,n_i} - p\right]_e dV_e + \left[\left(\frac{\partial(\Delta F_{mix})}{\partial T}\right)_{V,\varphi,n_i} - S_c\right]_e dT$$

$$+ \mu_{sl,e}dn_{sl,e} + \left(\sum_i \mu_i dn_i\right)_e + \left(\sum_i ez_i n_i d\varphi\right)_e \tag{S7}$$

where we use subscript "e" to denote the external solution. We consider the case of constant temperature, and note that $dV=-dV_o$, $dn_i=-dn_{i,o}$, and $dn_{sl}=-dn_{sl,o}$, and $dn_{pl}=0$. Thus, the total Helmholtz free energy of the combined system can be written as Eq. (7).

**Latent Heat of Evaporation.** We consider hydrogel is at equilibrium with saturated water vapor outside at pressure $p_s$. Inside hydrogel, water is at different pressure $p$. Using Eq. (23), we can write the water molar entropy as

$$s_w(T,p) = -\left(\frac{\partial \mu_w}{\partial T}\right)_{p,n_i}$$
$$= s_w^*(T,p) - R\left[\ln(1 - f_{2,eq}) + f_{2,eq} + \chi f_{2,eq}^2\right] - RTf_{2,eq}^2 \frac{\partial \chi}{\partial T} \tag{S8}$$

where $s_w^*(T,p)$ is the entropy of pure water at pressure T and p. The $\frac{\partial \chi}{\partial T}$ term arises because entropy caused by molecular configuration change around the contacting region between water and polymer molecules is also included in Eq.(S3). In an ideal mixing model,[10] this term actually cancels the $\chi f_2^2$ term in the square brackets so that only mixing of entropy term is left. The latent heat of evaporation is

$$L = h_{w,v}^*(T,p_s) - h_w(T,p) = T[s_{w,v}^*(T,p_s) - s_w(T,p)]$$

$$= L_o + RT[\ln(1-f_{2,eq}) + f_{2,eq} + \chi f_{2,eq}^2] + RT^2 f_{2,eq}^2 \frac{\partial \chi}{\partial T} \tag{S9}$$

where additional subscript "v" is used to represent the vapor phase, and $L_o = h_{w,v}^*(T,p_s) - h_w^*(T,p_s)$ is latent heat of evaporation of pure water at temperature T and $p_s$. In deriving the above expression, we took $s_w^*(T,p_s) - s_w^*(T,p) = 0$, which can be justified by the Maxwell relationship $(\partial s^*/\partial p)_T = -(\partial v^*/\partial T)_p$ by neglecting the thermal expansion of pure water.

**Freezing Point Depression and Boiling Point Elevation.** First, let's examine the melting point depression. We assume pure ice is formed inside the hydrogel, whose chemical potential can be written as

$$d\mu_{ice} = v_{ice}dp - s_{ice}dT + RTd\ln(a_{ice}) = v_{ice}dp - s_{ice}dT \tag{S10}$$

From Eq.(23), the change in the chemical potential of water can be written as

$$d\mu_w = v_w^* dp - s_w^* dT + RTd[\ln(1-f_2) + f_2 + \chi f_2^2] \tag{S11}$$

Note T in the last term on the right-hand side is outside the differentiation as the derivative is taken with (p,T) kept constant by definition. At freezing equilibrium, we have

$$v_{ice}dp - s_{ice}dT_{fz} = v_w^* dp - s_w^* dT_{fz} + RT_{fz}d[\ln(1-f_2) + f_2 + \chi f_2^2] \tag{S12}$$

where $T_{fz}$ represents the freezing point. For typical mixture analysis, the pressure does not change. However, in hydrogels, as $f_2$ increases from 0 (at which it is normal ice-water at the ambient pressure) to a finite value, the pressure inside the hydrogel increases. If we assume that ice forms inside hydrogel and experience same pressure as water, we should include the pressure change as $f_2$ increases. With the above argument, we get

$$(s_w^* - s_{ice})dT_{fz} = RT_{fz}d[\ln(1-f_2) + f_2 + \chi f_2^2] + (v_w^* - v_{ice})dp \tag{S13}$$

Replacing $(s_w^* - s_{ice})$ on the left hand side by $L_m/T_{fz}$, where $L_m$ is the latent heat of melting. We can also approximate $v_w^* \approx v_{ice}$ and neglect the last term. This approximation allows integration of Eq. (S12), leading to Eq. (26).

At the boiling point, the chemical potential equals the pure vapor chemical potential. We can follow similar steps as in the freezing point depression, replacing "ice" subscript in Eq. (S13) by "v" for vapor. We consider the boiling happens on surface of hydrogel so that the outside vapor phase is at a constant pressure while the liquid phase pressure inside the hydrogel depends on $f_2$. Equation (S13) becomes

$$-[s_{w,v}(T_{bp}, p_v) - s_w^*(T_{bp}, p)]dT_{bp} + v_{w,v}dp_v = RT_{bp}d[ln(1-f_2) + f_2 + \chi f_2^2] + v_w^* dp \quad (S14)$$

where $T_{bp}$ is the boiling point. The outside vapor pressure $p_v$ does not change during measurement. The first term again can be related to latent heat of evaporation (the entropy dependence on pressure for water is small). Using Eq. (13) for pressure change, we can write the above equation into

$$-\left[L_o/R + KT_{bp}\left(f_2^{1/3} - \frac{f_2}{2}\right)\right]\frac{dT_{bp}}{T_{bp}^2} = d[ln(1-f_{2,eq}) + f_{2,eq} + \chi f_{2,eq}^2] + K\, d\left(f_2^{1/3} - \frac{f_2}{2}\right) \quad (S15)$$

The second term inside the square bracket on the left-hand side arises from temperature dependence of pressure in Eq.(13), but its value is small relative to the latent heat of evaporation as we discussed. We can neglect this term and arrive at Eq. (27).

**Salt Content Inside Hydrogel in Equilibrium with Salty Water Outside.** Adding up Eqs. (30) and (31) and using charge neutrality $x_{Na+} = x_{Cl-}$, we get

$$\mu_{NaCl} = \mu_{NaCl}^o + v_{NaCl}(p - p_o) + 2RT[ln(1-f_2) + f_2 + \chi f_2^2] + RTln(\gamma_{NaCl}x_{NaCl})^2 \quad (S16)$$

where $\gamma_{NaCl} = \sqrt{\gamma_{Na+}\gamma_{Cl-}}$ is the activity coefficient of NaCl, and $v_{NaCl} = v_{Na+} + v_{Cl-}$ is the molar volume of NaCl. For water and ions outside hydrogel, we have

$$\mu_{w,e} = \mu_w^o + RTln(\gamma_w x_w)_e \quad (S17)$$

$$\mu_{Na+,e} = \mu_{Na+}^o + RTln(\gamma_{Na+}x_{Na+})_e \quad (S18)$$

$$\mu_{Cl-,e} = \mu_{Cl-}^o + RTln(\gamma_{Cl-}x_{Cl-})_e \quad (S19)$$

$$\mu_{NaCl,e} = \mu_{NaCl}^o + RTln[(\gamma_{NaCl}x_{NaCl})^2]_e \quad (S20)$$

From Eqs. (S16) and (S20), we get Eq. (32). From Eq.(S17) and Eq. (23), we get Eq. (33).

Also, in this case, a membrane potential might exist. To show this possibility, we start with the chemical potential balance of individual ion species:

$$v_{Na+}(p - p_o) + F\varphi + RT[ln(1-f_2) + f_2 + \chi f_2^2 + ln(\gamma_{Na+}x_{Na+})] = RTln(\gamma_{Na+}x_{Na+})_e \quad (S21)$$

$$v_{Cl-}(p - p_o) - F\varphi + RT[ln(1-f_2) + f_2 + \chi f_2^2 + ln(\gamma_{Cl-}x_{Cl-})] = RTln(\gamma_{Cl-}x_{Cl-})_e \quad (S22)$$

If the activity coefficients of the ions are equal, subtracting the above two equations leads to Eq. (34).

**Solubility of Salts inside Hydrogel.** From Eq. (S16), we have the chemical potential of NaCl in the solution as

$$d\mu_{NaCl} = d\mu^*_{NaCl}(p,T) + RTd\{ln(\gamma x_{NaCl,l})^2\} + 2RTd\{[ln(1-f_2) + f_2 + \chi f_2^2]\} \quad (S23)$$

when salt in solution is at equilibrium with solid salt, we have

$$d\mu^*_{NaCl,s}(p,T) = d\mu^*_{NaCl}(p,T) + RTd\{ln(\gamma x_{NaCl,l})^2\} + 2RTd\{[ln(1-f_2) + f_2 + \chi f_2^2]\} \quad (S24)$$

where $\mu^*_{NaCl,s}(p,T)$ is the chemical potential of pure solid salt. At the melting point ($T_m$=801 °C for NaCl), pure NaCl solid and liquid are at equilibrium with $x_{NaCl}$=1 and $f_2$=0. The chemical potentials of pure substance can be similarly expressed by entropy as in Eqs. (S10) and (S11), and an equation like Eq. (S13) can be integrated for temperature to change from $T_m$ to $T$, $x_{NaCl}$ from 1 to its solubility in water, and $f_2$ from 0 to a given value. Since the temperature range is large, the entropy change with temperature for both the solid and liquid phase may need to be included. Instead of direct integration of Eq. (S24), it is easier to use the Gibbs-Helmholtz

$$\left(\frac{\partial(\mu_i/T)}{\partial T}\right)_p = -\frac{h_i}{T^2} \quad (S25)$$

and write down the chemical potential change for pure NaCl in liquid and solid phase as

$$\frac{\mu^*_{NaCl,s}(T,p)}{T} - \frac{\mu^*_{NaCl,s}(T_m,p)}{T_m} = H_s(T_m,p)\left(\frac{1}{T} - \frac{1}{T_m}\right) - \int_{T_m}^{T}\left[\frac{1}{T^2}\int_{T_m}^{T}c_{p,sd}dT\right]dT \quad (S26)$$

$$\frac{\mu^*_{NaCl}(T,p)}{T} - \frac{\mu^*_{NaCl}(T_m,p)}{T_m} = H_l(T_m,p)\left(\frac{1}{T} - \frac{1}{T_m}\right) - \int_{T_m}^{T}\left[\frac{1}{T^2}\int_{T_m}^{T}c_{p,lq}dT\right]dT \quad (S27)$$

The above relation leads to

$$\mu^*_s(T,p) - \mu^*_l(T,p) = -L_{sl}\left(1 - \frac{T}{T_m}\right) + T\int_{T_m}^{T}\left[\frac{1}{T^2}\int_{T_m}^{T}(c_{p,l} - c_{p,s})dT\right]dT \quad (S28)$$

where $L_m$ is the latent heat of the solid-liquid phase transition at $T_m$ and $p$. We will neglect the specific heat term, as is often done in literature. This can be justified because the specific heat difference between liquid and solid NaCl is ~50 J/kg-K,[63] while the latent heat is 1460 kJ/kg. In this case, setting the chemical of the solid and liquid phases equaling each other, and using Eq. (S28), we get Eq. (36).

**Polyelectrolyte Hydrogel in Equilibrium with Salty Water.** Balancing the chemical potential for each mobile species, we arrive at

$$K\left[f_2^{1/3} - \frac{f_2}{2}\right] + [ln(1 - f_2) + f_2 + \chi f_2^2 + ln(\gamma_w x_w)] = ln(\gamma_w x_w)_e \qquad (S29)$$

$$K_{Na+}\left[f_2^{1/3} - \frac{f_2}{2}\right] + \frac{F\varphi}{RT} + [ln(1 - f_2) + f_2 + \chi f_2^2 + ln\gamma_{Na+}x_{Na+}] = ln(\gamma_{Na+}x_{Na+})_e \qquad (S30)$$

$$K_{Cl-}\left[f_2^{1/3} - \frac{f_2}{2}\right] - \frac{F\varphi}{RT} + [ln(1 - f_2) + f_2 + \chi f_2^2 + ln\gamma_{Cl-}x_{Cl-}] = ln(\gamma_{Cl-}x_{Cl-})_e \qquad (S31)$$

with the condition

$$x_{Na+} + x_{Cl-} + x_w = 1 \qquad (S32)$$

Solving Eqs.(S29)-(S32) will give us the equilibrium volume fraction $f_2$, concentrations of Na$^+$ and Cl$^-$ ions and water, and the Donnan potential $\varphi$. Examples of the solution are given in Fig.9.

**Reference**

[63] C.-J. Li, P.W. Li, K. Wang, E.E. Molina, AIMS Energy, 2014, 2, 133-157.